\documentclass[aps,prc,preprint,amsmath,amssymb,showpacs,preprintnumbers,superscriptaddress]{revtex4-1}
\usepackage{CJK}
\usepackage{graphicx}
\usepackage{dcolumn}
\usepackage{bm}
\usepackage{color}
\usepackage{hyperref}

\allowdisplaybreaks

\begin{document}
\begin{CJK*}{GB}{}
\title{Entanglement in multinucleon transfer reactions}
\CJKfamily{gbsn}
\author{B. Li \CJKfamily{gbsn}(À)}
\affiliation{State Key Laboratory of Nuclear Physics and Technology, School of Physics, Peking University, Beijing 100871, China}
\author{D. Vretenar}
\email{vretenar@phy.hr}
\affiliation{Physics Department, Faculty of Science, University of Zagreb, 10000 Zagreb, Croatia}
\affiliation{State Key Laboratory of Nuclear Physics and Technology, School of Physics, Peking University, Beijing 100871, China}
\author{T. Nik\v si\' c}
\affiliation{Physics Department, Faculty of Science, University of Zagreb, 10000 Zagreb, Croatia}
\affiliation{State Key Laboratory of Nuclear Physics and Technology, School of Physics, Peking University, Beijing 100871, China}
\author{D. D. Zhang \CJKfamily{gbsn}(Õŵ¤µ¤)}
\affiliation{State Key Laboratory of Nuclear Physics and Technology, School of Physics, Peking University, Beijing 100871, China}
\author{P. W. Zhao \CJKfamily{gbsn}(ÕÔÅôΡ)}
\email{pwzhao@pku.edu.cn}
\affiliation{State Key Laboratory of Nuclear Physics and Technology, School of Physics, Peking University, Beijing 100871, China}
\author{J. Meng \CJKfamily{gbsn}(ÃϽÜ)}
\email{mengj@pku.edu.cn}
\affiliation{State Key Laboratory of Nuclear Physics and Technology, School of Physics, Peking University, Beijing 100871, China}

\begin{abstract}
Nuclear reactions present an interesting case for studies of the time-evolution of entanglement between complex quantum systems. In this work, the time-dependent nuclear density functional theory is employed to explore entanglement in multinucleon transfer reactions. As an illustrative example, for the reaction $^{40}$Ca $+$ $^{208}$Pb at $E_{\rm lab} = 249$ MeV, in the interval of impact parameters $4.65-7.40$ fm, and the relativistic density functional PC-PK1, we compute the von Neumann entropies, entanglement between fragments, nucleon-number fluctuations, and Shannon entropy for the nucleon-number observable. A simple linear correlation is established between the entanglement and nucleon-number fluctuation of the final fragments. The entanglement between the fragments can be related to the corresponding excitation energies and angular momenta. The relationship between the von Neumann entropy and the Shannon entropy for the nucleon-number observable is analyzed, as well as the time-evolution of the entanglement (nucleon-number fluctuation). The entanglement is also calculated for a range of incident energies and it is shown how, depending on the impact parameter, the entanglement increases with the collision energy.
\end{abstract}

\date{\today}

\maketitle

\section{Introduction}
The phenomenon of quantum entanglement, where microscopic systems become interconnected and exhibit correlated behavior regardless of the distance between them, has been explored and applied in many fields of physics
\cite{Amico2008RMP,Horodecki2009RMP,Laflorencie2016PR,Deng2017PRX,Friis2019NRP,Abanin2019RMP,Erhard2020NRP,Lu2022RMP}.
One of the more recent is low-energy nuclear physics. Even though it is obvious that an atomic nucleus is the epitome of a finite complex quantum system with entangled constituents, it is far from clear how the level of entanglement is manifested in various observables that characterize nuclear structure, reactions, and decay properties  \cite{Legeza2015PRC,Beane2019PRL,Robin2021PRC,Faba2021PRA,Kruppa2021JPG,Bulgac2023aPRC,Bulgac2023PRC,Pazy2023PRC,Sun2023PRC,Gu2023PRC,Johnson2023JPG,Lacroix2024}. Studies of the role of entanglement provide additional insights into the interactions that govern the behavior of nuclei, and the way nucleonic matter is organized in finite nuclear systems. Most applications have so far been focused on static nuclear structure properties, although implications for heavy-ion collisions and nuclear fission have also been discussed. In this work we explore entanglement in multinucleon transfer reactions.

Multinucleon transfer (MNT) reactions provide a unique and detailed perspective on nuclear structure and dynamics, and hold promise for the synthesis of yet-unknown neutron-rich nuclei and superheavy nuclei \cite{Dasso1994PRL,Zagrebaev2005JPG,Zagrebaev2006PRC,Zagrebaev2007JPG,Zagrebaev2008PRL,Zagrebaev2008PRC,Zagrebaev2011PRC,Zagrebaev2013PRC,Watanabe2015PRL}. Data obtained in
MNT reactions play an important role for studies of shell evolution \cite{Otsuka2020RMP}, shape transition \cite{Cejnar2010RMP,Heyde2011RMP}, and r-process nucleosynthesis \cite{Cowan2021RMP}.
Compared to fragmentation, fission, and fusion processes, which have successfully been applied to extend the landscape of known isotopes, MNT reactions are more likely to produce unstable neutron-rich superheavy isotopes located in the north-eastern part of the chart of nuclides \cite{Watanabe2015PRL,Zagrebaev2008PRL,Adamian2020EPJA}.
In addition, they can be employed for detailed investigations of the interaction between medium-heavy and heavy nuclei, and the mechanism of nucleon exchange during a nuclear collision. As time-dependent processes, MNTs are ideal for studies of the time-evolution of entanglement between strongly interacting complex systems.

Because of the importance of MNT reactions and the wealth of available data,
numerous theoretical models have been developed and applied to describe reaction dynamics,
including both phenomenological and microscopic approaches.
Among the latter, nuclear density functional theory (DFT) very successfully reproduces and predicts structure properties across the chart of nuclides
\cite{Vautherin1972PRC,Decharge1980PRC,Ring1996PPNP,Vretenar2005PR,Niksic2011PPNP,Meng2016}.
Its dynamic extension, the time-dependent density functional theory (TD-DFT), presents a general microscopic framework that, without any parameter specifically adjusted to the reaction mechanism, can be employed to analyze MNT reactions \cite{Simenel2010PRL,Sekizawa2013PRC,Sekizawa2014PRC,Sekizawa2016PRC,Sekizawa2017PRCR,Sekizawa2017PRC,Sekizawa2019FIP,Wu2019PRC,Jiang2020PRC,Wu2022PLB}.
Connections between the reaction products and the entrance channel characteristics, such as the neutron-to-proton (N/Z) ratio \cite{Sekizawa2013PRC},
the charge product $Z_{P}Z_{T}$ \cite{Sekizawa2013PRC}, and the relative orientation of deformed ions \cite{Sekizawa2013PRC,Sekizawa2016PRC,Sekizawa2017PRCR,Wu2019PRC},
have been explored in TD-DFT studies.
The time-dependent covariant density functional theory (TD-CDFT), as the relativistic extension of TD-DFT,
has been developed \cite{Ren2020PLB,Ren2020PRC} and successfully applied to various nuclear processes,
including $\alpha$ cluster scattering \cite{Ren2020PLB}, fusion \cite{Ren2020PRC}, chirality \cite{Ren2022b},
fission \cite{Ren2022PRL,Ren2022PRC,Li2023PRC_FT,Li2023PRC_Gd,Li2024FOP}, quasifission \cite{Zhang2024PRC} and, very recently, MNT reactions \cite{Zhang2024PRC_MNT}.

In this work, TD-CDFT is applied to a study of entanglement between fragments, and information entropy in MNT reactions.
Section~\ref{sec_theo} outlines the TD-CDFT formalism,
and the methods that will be used to compute the entanglement between fragments, excitation energies, angular momenta, and the Shannon entropy for the nucleon-number observable.
Numerical details for the self-consistent static and dynamic calculations are explained in Sec.~\ref{sec_num}.
In Sec.~\ref{CaPb}, we present and discuss the results of a study of entanglement in the MNT reaction $^{40}$Ca $+$ $^{208}$Pb.
Finally, Sec.~\ref{sec_sum} concludes the article with a summary and brief outlook for future studies.
\section{Theory framework}\label{sec_theo}
\subsection{Time-dependent covariant density functional theory}
 In the framework of time-dependent (covariant) density functional theory (TD-CDFT)~\cite{Ren2020PRC, Ren2020PLB}, the nuclear wave function is at all times a Slater determinant of occupied single-particle states,
\begin{equation}
   |\Psi(t)\rangle = \prod_{k=1}^A c_{k}^\dagger(t)|-\rangle,
\label{Eq_Slater}
\end{equation}
where $c_{k}^\dagger(t)$ is the time-dependent creation operator for the single-nucleon wave functions $\psi_k(\bm{r},t)$, and $A$ is the total number of nucleons.
The evolution of single-nucleon wave functions is governed by the time-dependent Dirac equation,
\begin{equation}\label{Eq_td_Dirac_eq}
  i\frac{\partial}{\partial t}\psi_k(\bm{r},t)=\hat{h}(\bm{r},t)\psi_k(\bm{r},t),
\end{equation}
where the single-particle Hamiltonian $\hat{h}(\bm{r},t)$ reads
\begin{equation}
   \hat{h}(\bm{r},t) = \bm{\alpha}\cdot(\hat{\bm{p}}-\bm{V}(\bm{r},t))+V^0(\bm{r},t)+\beta(m+S(\bm{r},t)).
   \label{Ham_D}
\end{equation}
$m$ is the nucleon mass, and
the scalar $S(\bm{r},t)$ and four-vector $V^{\mu}(\bm{r},t)$ potentials are self-consistently determined at each step in time
by the instantaneous densities and currents in the isoscalar-scalar, isoscalar-vector and isovector-vector channels
\begin{subequations}\label{Eq_density_current}
  \begin{align}
    &\rho_S(\bm{r},t)=\sum_{k=1}^A\bar{\psi}_k(\bm{r},t)\psi_k(\bm{r},t),\\
    &j^\mu(\bm{r},t)=\sum_{k=1}^A\bar{\psi}_k(\bm{r},t)\gamma^\mu\psi_k(\bm{r},t),\\
    &j_{TV}^\mu(\bm{r},t)=\sum_{k=1}^A\bar{\psi}_k(\bm{r},t)\gamma^\mu\tau_3\psi_k(\bm{r},t).
  \end{align}
\end{subequations}

In the present analysis, we will consider a collision of two closed-shell nuclei described by TD-CDFT.
Before the collision, the projectile and target are composed of $A_P^i$ and $A_T^i$ nucleons, respectively, and the wave function of the system is a single Slater determinant with the total number of nucleons $A=A_T^i+A_P^i$.
The whole 3D-lattice space can be divided into two subspaces, i.e., the projectile subspace $V^i_P$ and target subspace $V^i_T$. For a choice of initial conditions, that is, the projectile kinetic energy and impact parameter, TD-CDFT models the most probable path of collision dynamics. After the collision, it is assumed that two nuclei emerge, a projectile-like fragment (PLF) and a target-like fragment (TLF), and the whole space can again be divided into two subspaces, i.e., the projectile-like subspace $V^f_P$ and target-like subspace $V^f_T$.
We note that the wave functions of the PLF and TLF are generally not eigenstates of the particle number operator
but a superposition of states with different nucleon numbers.
The average nucleon number $\bar{A}_P^f$ of PLF and the average nucleon number $\bar{A}_T^f$ of TLF satisfy the particle number conservation relation $A=\bar{A}_P^f+\bar{A}_T^f$.
The interface between the projectile-like subspace $V^f_P$ and target-like subspace $V^f_T$ is well-defined when the two emerging fragments are sufficiently separated.
During the collision stage in which the projectile and target partially overlap, and if a di-nuclear structure can be discerned,
it is also possible to define an interface in the neck region and divide the whole space into a PLF subspace and a TLF subspace (See Fig.1 in Ref.~\cite{Sekizawa2020PRC}).
First, by diagonalizing the mass quadrupole matrix as described in Refs.~\cite{Ayik2018PRC,Yilmaz2014PRC}, one determines the elongation axis of the total system, which is perpendicular to the interface between the PLF and TLF subspaces. Then, one can place the interface at the minimum density (in the neck region) location along the elongation axis, as in Refs.~\cite{Ayik2015PRC,Yilmaz2014PRC,Ayik2018PRC,Ayik2019PRC}. In this way, the whole space is divided into the subspace $V$, which contains the fragment we are interested in, and the complementary subspace $\bar{V}$, at each time $t$. We also note that alternative methods can be used to assign particles to each fragment during the phase of partial overlap. For instance, a quantum localization method based on the partition of orbital wave functions into two sets belonging to the two emerging fragments \cite{Younes2011PRL}.

\subsection{Entropy of fragments and entanglement}
The von Neumann entropy is defined in terms of the density matrix
\begin{equation}
S = - {\rm Tr} (\rho \ln \rho).
\end{equation}
For a pure state, $S=0$. In the case of separate fragments, it is defined in terms of the reduced density matrices $\rho_{\rm \sc PLF} = {\rm Tr}_{\rm \sc TLF} (\rho)$ and $\rho_{\rm \sc TLF} = {\rm Tr}_{\rm \sc PLF} (\rho)$, and the traces are over the TLF and PLF, respectively.
The von Neumann entropy $S^{(q)}_V$ for neutrons ($q=n$) or protons $(q=p)$ of the fragment, located in the subspace $V$, can be obtained by the method introduced in Ref.~\cite{Klich2006JPA},
\begin{equation}
\begin{aligned}
   S_V^{(q)}&=-{\rm Tr} \{ {\bm M_V^{(q)}} \ln  {\bm M_V^{(q)}} + [{\bm I}- {\bm M_V^{(q)}}] \ln [{\bm I}- {\bm M_V^{(q)}}]\}\\
         &=-\sum_{i=1}^{N^{(q)}} \{ d_i^{(q)} \ln d_i^{(q)} + [1-d_i^{(q)}] \ln [1-d_i^{(q)}]\},
\end{aligned}
   \label{SV}
\end{equation}
where ${\bm I}$ is a unit matrix and $N^{(q)}$ is the number of neutrons ($q=n$) or protons $(q=p)$ of the total system.
$d_i$ are the eigenvalues of the matrix $ {\bm M_V^{(q)}}$, whose elements are defined by the relation
\begin{equation}
   [{\bm M_V^{(q)}}]_{ij} = \langle \psi_i^{(q)} |\hat{\Theta}_V|\psi_j^{(q)}\rangle,
   \label{MV}
\end{equation}
where $\hat{\Theta}_V$ is the Heaviside function in coordinate space
\begin{equation}
\begin{split}
    \Theta_V(\bm r)=\left \{
\begin{array}{ll}
    1,     & {\rm if}~~\bm{r} \in V\\
    0,     & {\rm if}~~\bm{r} \not\in V
\end{array}
\right.
\end{split}.
\end{equation}
The mean-value of the nucleon number $N^{(q)}_V$, and its fluctuation $\Delta N^{(q)2}_V$ can also be obtained from the matrix $ {\bm M_V^{(q)}}$~\cite{Klich2006JPA},
\begin{equation}
N^{(q)}_V={\rm Tr} [ {\bm M_V^{(q)}} ]=\sum_{i=1}^{N^{(q)}} d_i^{(q)},
\label{particle_numebr}
\end{equation}
\begin{equation}
\Delta N^{(q)2}_V={\rm Tr}\{ {\bm M_V^{(q)}}[{\bm I}- {\bm M_V^{(q)}}] \}=\sum_{i=1}^{N^{(q)}}  d_i^{(q)}[1-d_i^{(q)}].
\label{particle_fluctuation}
\end{equation}
The entropy of the fragment in the subspace $V$ is the sum of its proton and neutron entropies,
\begin{equation}
S_V=S_V^{(n)}+S_V^{(p)}.
\end{equation}

Using the same method, one can obtain the entropy $S_{\bar{V}}$ of the fragment in the complementary subspace $\bar{V}$. The corresponding matrix $ {\bm M_{\bar{V}}^{(q)}}$ satisfies the equation ${\bm M_{\bar{V}}^{(q)}}={\bm I}- {\bm M_V^{(q)}}$, and, thus, $S_{\bar{V}}=S_{V}$. If we start from a pure state, that is, a product state of a projectile and target with well defined nucleon numbers $A_P^i$ and $A_T^i$, the von Neumann entropy of the total system is zero at all times because the time-evolution is unitary. $S_V$ and $S_{\bar{V}}$, however, correspond to the reduced density operators, and these entropies are not time-independent. The reduced density matrices have the same eigenvalues, and the entropy and nucleon number fluctuation of the fragment in the complementary subspace $\bar{V}$, are same as the ones in the subspace $V$. The entanglement (mutual information) between the fragment in the subspace $V$, and the fragment in subspace $\bar{V}$, can be defined \cite{Auletta2009Book}
\begin{equation}
L=S_V+S_{\bar{V}}-S_{tot},
\label{entanglement}
\end{equation}
where $S_{\rm tot}$ is the entropy of total system, and it is computed by extending $V$ to the whole 3D-space. In the present case $S_{\rm tot}\equiv 0$, while $S_V$ and $S_{\bar{V}}$ are identical, and greater than zero if the total wave function is an entangled state. Since the time-evolution of the reduced density operators is not unitary, the entropies $S_V$ and $S_{\bar{V}}$ are time-dependent.

\subsection{Fragment excitation energy and angular momentum}
The total excitation energy of the system can be evaluated from the energy conservation relation
\begin{equation}
E^{*}_{\rm tot}=E^{\rm init}-E^{\rm TKE}-Q,
\end{equation}
where $E^{\rm init}$ is the initial energy, $E^{\rm TKE}$ is the average total kinetic energy in center-of-mass frame after the fragments separate, and $Q$ denotes the average $Q$-value of the particular reaction channel.

The average total kinetic energy of the outgoing fragments is defined by
\begin{equation}
 E^{\rm TKE}=\frac{1}{2}m\bar{A}^f_P |{\bm{v}_P^{f}}|^2+\frac{1}{2}m\bar{A}^f_T |{\bm{v}_T^{f}}|^2+E_{\rm Coul},
\end{equation}
where the velocity of the fragment $l=P,~T$ reads
\begin{equation}
{\bm v_l^f}=\frac{1}{m\bar{A}_l^f}\int_{V_l^f} d \bm{r}~\bm{j} (\bm{r}),
\end{equation}
and $\bm{j} (\bm{r})$ is the total current density.
The integration is over the subspace $V_l^f$ corresponding to the fragment $l$, and $E_{\rm Coul}$ is the Coulomb energy.

The average $Q$-value is evaluated from
\begin{equation}
    Q=M^{P,f}_{\bar{N}_P^f,\bar{Z}_P^f}+M^{T,f}_{\bar{N}_T^f,\bar{Z}_T^f}-M^{P,i}-M^{T,i},
\end{equation}
where $M^{l,i}$ is the mass of the initial nucleus $l=P,~T$.
The quantity $M^{l,f}_{\bar{N}_l^f,\bar{Z}_l^f}$ is the mass of fragment $l$, with $\bar{N}_l^f$ and $\bar{Z}_l^f$ the corresponding average neutron number and proton number, respectively.
$M^{l,f}_{\bar{N}_l^f,\bar{Z}_l^f}$ can be evaluated by linear interpolation from the masses of its neighboring nuclei. To evaluate the average $Q$-values, experimental masses from the latest atomic mass evaluation,  AME2020~\cite{Huang2021CPC,Wang2021CPC}, have been used. In a first approximation, one may distribute the total excitation energy to the respective fragments in such a way that it is proportional to their masses~\cite{Sekizawa2017PRC},
\begin{equation}
E^{*}_{l}=\frac{\bar{A}_l^f}{A}E^{*}_{\rm tot}=\frac{\bar{N}_l^f+\bar{Z}_l^f}{A}E^{*}_{\rm tot}.
\end{equation}

The average angular momentum of fragment $l$ is determined by
\begin{equation}
\bm{J}_{l}=\langle \Psi| \hat{\bm J}_{V_l^f}|\Psi\rangle=\langle \Psi| \sum_{k=1}^{A}  \hat{\Theta}_{V_l^f}\hat{\bm j}_k|\Psi\rangle,
\end{equation}
where $\hat{\bm j}_k=(\hat{\bm r}_k-{\bm R}_{\rm c.m.})\times \hat{\bm p}_k+\hat{\bm s}_k$.
${\bm R}_{\rm c.m.}$ is the center-of-mass coordinate of the total system,
and $\hat{\bm p}_k$ and $\hat{\bm s}_k$ are the single-particle momentum and spin operators, respectively.

\subsection{Shannon entropy of fragments}
The measurement entropy of an observable $\hat O$ can be expressed in terms of the Shannon entropy
\begin{equation}
    H [ \hat O ]=-\sum_{x}P({x}) \ln P(x),
\end{equation}
where $P({x})$ is the probability distribution of the outcomes $x$ of $\hat O$. In particular, for the observable of the number of nucleons in a fragment after collision, the Shannon entropy~\cite{Eisert2010RMP} of the fragment in the subspace $V$ can be evaluated from
\begin{equation}
    H=-\sum_{N,Z}P_{N,Z} \ln P_{N,Z},
    \label{Shannon_entropy}
\end{equation}
where $P_{N,Z}$ is the probability of the occurrence in $V$ of a fragment composed of $N$ neutrons and $Z$ protons.
$P_{N,Z}$ can be computed in a standard way by employing particle number projection~\cite{Simenel2010PRL,Sekizawa2013PRC},
\begin{equation}
    P_{N,Z}=\langle\Psi|\hat{P}_N^{(n)}\hat{P}_Z^{(p)}|\Psi\rangle=P_N^{(n)}P_Z^{(p)},
    \label{Eq_P_NZ}
\end{equation}
where the particle number projection operator for neutrons $(q=n)$ or protons $(q=p)$ reads
\begin{equation}
\hat{P}_m^{(q)}=\frac{1}{2\pi}\int_0^{2\pi} d\theta~e^{i(m-\hat{N}_V^{(q)})\theta}.
\end{equation}
$\hat{N}_V^{(q)}$ is the particle number operator in the subspace $V$, defined as
\begin{equation}
\hat{N}_V^{(q)}=\int_V d{\bm r} \sum_{k=1}^{N^{(q)}} \delta(\bm{r}-\bm{r_k})=\sum_{k=1}^{N^{(q)}} \Theta_V(\bm{r}_k).
\end{equation}
$P_N^{(n)}$ and $P_Z^{(p)}$ are individual probabilities for $N$ neutrons and $Z$ protons, respectively, that are computed  from
\begin{equation}
P_m^{(q)}=\frac{1}{2\pi} \int_0^{2\pi} d\theta~e^{im\theta}\det \mathcal{B}^{(q)} (\theta),
\end{equation}
where
\begin{equation}
[\mathcal{B}^{(q)} (\theta)]_{ij}=\langle \psi_i^{(q)}| \psi_j^{(q)} (\theta)\rangle,
\end{equation}
and
\begin{equation}
    \psi_j^{(q)}(\bm{r},\theta)=[1+(e^{-i\theta}-1)\Theta_V(\bm{r})]\psi_j^{(q)}(\bm{r}).
\end{equation}
The number of protons and number of neutrons are independent observables and, thus, the Shannon entropy can be written as the sum of two terms
\begin{equation}
     H=-\sum_{N,Z}P_{N,Z} \ln P_{N,Z}=-\sum_N P^{(n)}_N \ln P^{(n)}_N - \sum_Z P^{(p)}_Z \ln P^{(p)}_Z=H^{(n)}+H^{(p)},
\end{equation}
where $H^{(n)}$ and $H^{(p)}$ are the Shannon entropies of neutrons and protons, respectively, and it is assumed that the nucleon number probability distributions are normalized.
\section{Numerical details}\label{sec_num}
In this work, we consider the multinucleon transfer reaction $^{40}$Ca+$^{208}$Pb and analyze the von Neumann entropies, entanglement, and Shannon entropies, as functions of the impact parameter and collision energy.
The point-coupling relativistic energy density functional PC-PK1~\cite{Zhao2010PRC} is used both for static and dynamic calculations.
Before the collision, the projectile and target are in their ground states determined by self-consistent relativistic DFT calculations in a three-dimensional lattice space, using the inverse Hamiltonian and Fourier spectral methods \cite{Ren2017PRC,Ren2019SCPMA,Ren2020NPA}, with the box size: $L_x\times L_y\times L_z=20.8\times20.8\times20.8~{\rm fm}^3$.
In calculations with TD-CDFT, the mesh spacing of the lattice is 0.8 fm for all directions, and the box size is $L_x\times L_y\times L_z=48\times20.8\times48~{\rm fm}^3$.
The time-dependent Dirac equation \eqref{Eq_td_Dirac_eq} is integrated using the predictor-corrector method.
The step for the time-evolution is $0.2~{\rm fm}/c$.
At the initial time, the projectile and target are placed on the mesh at a distance of $20$ fm between them,
and it is assumed they initially follow a Rutherford trajectory.
After the collision, if two distinct fragments are produced, the time-evolution is completed
when the distance between the fragments is larger than $20$ fm or $24$ fm, depending on the impact parameter.

\section{Results and discussion}\label{CaPb}
\begin{figure}[!htbp]
\centering
\includegraphics[width=0.4\textwidth]{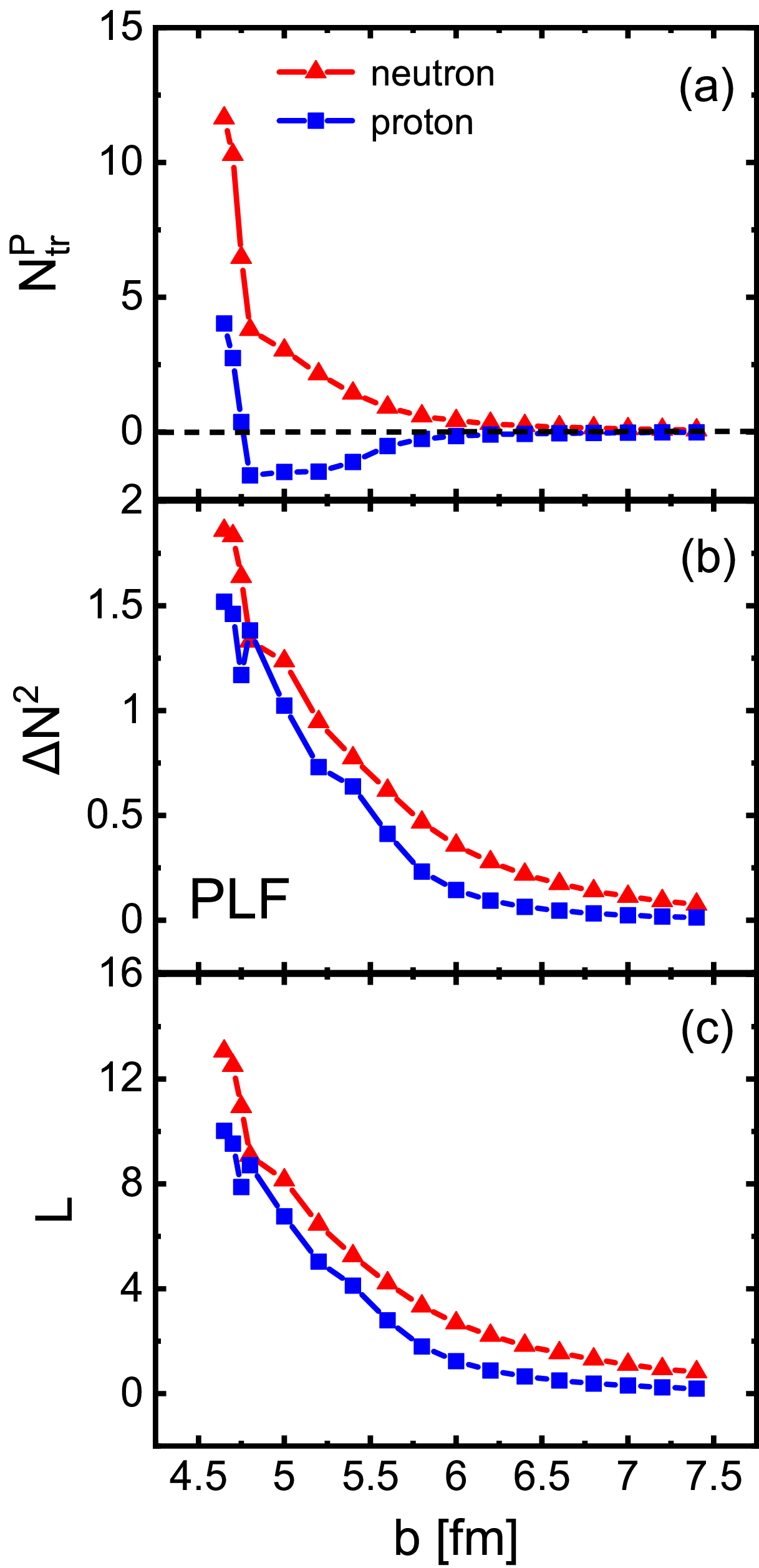}
\caption{(Color online) $^{40}$Ca+$^{208}$Pb reaction at $E_{\rm lab}=249$ MeV. (a) Average number of transferred nucleons from the target to the projectile.
(b) Nucleon-number fluctuations of the projectile-like fragment (PLF). (c) The entanglement between the projectile-like and target-like fragments. On the horizontal axis is the impact parameter $b$.
}
 \label{fig:entanglement}
\end{figure}

The self-consistent solutions for the ground states of $^{40}$Ca and $^{208}$Pb, based on the functional PC-PK1, are spherical.
The nuclei are initially boosted by a Lorentz transformation that corresponds to $E_{\rm lab}=249$ MeV, and the TD-CDFT calculation of the $^{40}$Ca+$^{208}$Pb reaction is carried out in the center-of-mass frame.
In panel (a) of Fig.~\ref{fig:entanglement}, we display the average number of transferred nucleons from the target ($^{208}$Pb) to the projectile ($^{40}$Ca), as a function of the initial impact parameter $b$.
As shown in our recent study \cite{Zhang2024PRC_MNT}, the largest value of the impact parameter for complete fusion in this reaction is $b = 4.63$ fm, when calculations are performed using the functional PC-PK1. For multi-nucleon transfer, therefore, we will consider the interval of impact parameters: $4.65-7.40$ fm, and analyze the two fragments, PLF and TLF,  that emerge from the collision.
The neutron-to-proton ratio of the projectile is 1, while that of the target is 1.54. One expects that, for collisions between nuclei with different N/Z values, the dominant transfer process is towards charge equilibrium, that is, nucleon transfer tends to equalize the N/Z ratio in the PLF and TLF. Here, this means neutron transfer from $^{208}$Pb to $^{40}$Ca, and proton transfer from $^{40}$Ca to $^{208}$Pb. The average number of transferred neutrons, denoted by red triangles, decreases at large impact parameters.
The average number of transferred protons, indicated by blue squares, exhibits a minimum at $b \approx 5$ fm.
These values are consistent with the results obtained in the non-relativistic TD-DFT framework~\cite{Sekizawa2013PRC}.
The sharp increase of both transferred neutrons and protons from the target to the projectile for $b < 5$ fm, reflects the fact that the reaction is approaching complete fusion. The corresponding particle-number fluctuations of the PLF are shown in
panel (b) of Fig.~\ref{fig:entanglement}. In general, as the interaction between the projectile and target decreases at large impact parameters, so does the nucleon-number fluctuation. The discontinuities at $b= 4.8$ fm, more pronounced for the proton-number fluctuation, are caused by the sudden increase of the number of transferred neutrons and protons for $b < 5$ fm.

\begin{figure}[!htbp]
\centering
\includegraphics[width=0.5\textwidth]{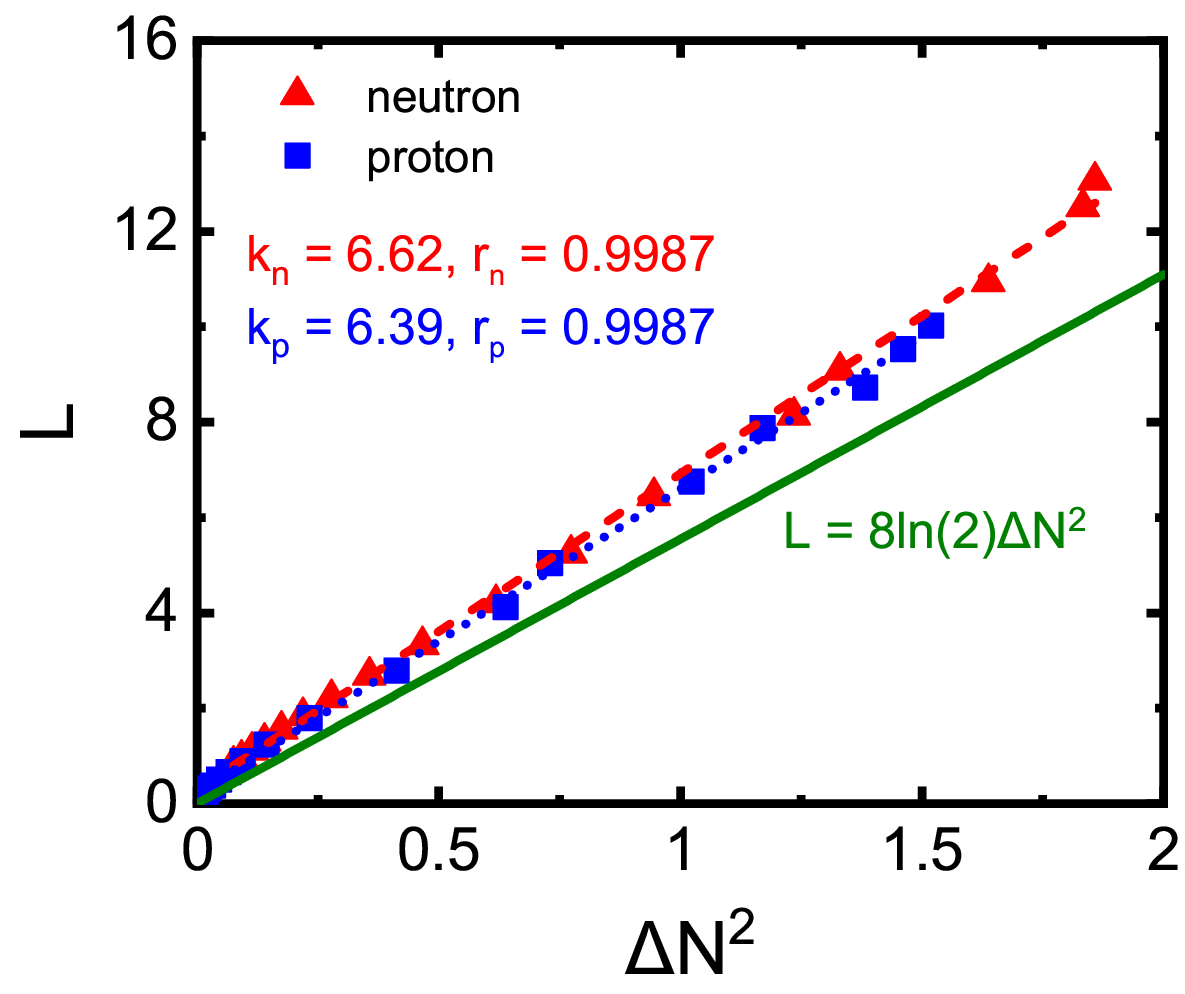}
\caption{(Color online) Linear regression model for the entanglement and particle-number fluctuations of the PLF for neutrons (red) or protons (blue), in the reaction $^{40}$Ca+$^{208}$Pb at $E_{\rm lab}=249$ MeV. $k_n$ ($k_p$) denotes the regression slope, and $r_n$ ($r_p$) is the correlation coefficient. The solid line corresponds to the expression
$L=(8\ln 2)~\Delta N^2$ \cite{Klich2006JPA}.}
 \label{fig:linear_regression}
\end{figure}
Since the wave function of the total system is a pure state, its von Neumann entropy vanishes, and the entanglement $L$ between the PLF and TLF, shown in panel (c) of Fig.~\ref{fig:entanglement}, is simply twice the von Neumann entropy of one of the fragments. It exhibits the same dependence on the value of the impact parameter as the nucleon-number fluctuations. In fact, as shown by the simple linear regression model in Fig.~\ref{fig:linear_regression}, there is a perfect linear relation between the entanglement and particle-number fluctuation, with the corresponding correlation coefficient $r = 0.9987$, both for neutrons and protons. The strong correlation between entanglement and particle-number fluctuations has also been pointed out in previous studies (cf. Ref.~\cite{Laflorencie2016PR} and references therein, as well as the recent analysis of entanglement entropy of subsets of qubits \cite{Lacroix2024}). A linear relation between entanglement and nucleon-number fluctuations has also recently been found in Ref.~\cite{Gu2023PRC}, which reported a study of entanglement entropies between single-particle states of the reference state (the hole space) and its complement (the particle space) in finite nuclei. In Ref.~\cite{Klich2006JPA} it has been demonstrated that, given a partition of the single-particle Hilbert space into orthogonal subspaces, the von Neumann entropy and nucleon-number fluctuation of a fragment in a region $V$ should satisfy the inequality
\begin{equation}
     S_V^{(q)} \geq (4\ln 2)~\Delta N^{(q) 2}_V.
     \label{inequility}
\end{equation}
In the present case the entanglement is twice the entropy of the PLF, so there should be an additional factor 2 on the right-hand side of the inequality. In fact, as shown in Fig.~\ref{fig:linear_regression}, the computed entanglement for protons and neutrons, as a function of the nucleon-number fluctuation, is always greater than $8\ln (2)~\Delta N^{(q) 2}_V$. The importance of this inequality is that it provides a lower bound for the entanglement entropy, expressed in terms of particle-number fluctuations that are, in principle, measurable \cite{Cook2023NaC}.

In Ref.~\cite{Cook2023NaC}, multiple identities that colliding heavy nuclei take on the path to fusion were investigated in a multi-nucleon transfer experiment. For the reaction $^{40}$Ca+$^{208}$Pb, the distributions in mass (A), atomic number (Z), and kinetic energy were measured at 12 center-of-mass energies, from 20\% to 1\% below the fusion barrier. A rapid change in the identities of nuclei was observed already outside the capture barrier. For the present study, of particular interest is Fig. 1 of the supplement to Ref.~\cite{Cook2023NaC}, which shows the Z and N distribution of reflected nuclei produced in reactions $^{40}$Ca+$^{208}$Pb at laboratory angle $\theta_{\rm lab}= 115^{\rm o}$ and center-of-mass energies from $E_{\rm c.m.} /V_B = 0.80$ to 0.99, where $V_B$ is the barrier height. As the collision energy increases, the Z and N distributions widen significantly, that is, the nucleon-number fluctuations exhibit a pronounced increase, and approach the line of isospin asymmetry equal to that of the compound nucleus $^{248}$No ($N/Z = 1.43$).

We have employed the TD-CDFT to model these reactions and compute the proton and neutron distributions of the PLF. In the  experiment, the reflected nuclei were measured at the laboratory angle $\theta_{\rm lab}= 115^{\rm o}$ for 12 center-of-mass energies $E_{\rm c.m.}=156.7-191.7$ MeV. In our calculation, the initial center-of-mass energies are the same as in the experiments, and the initial impact parameters are determined by the Rutherford scattering formula for $\theta_{\rm lab}= 115^{\rm o}$. For very small impact parameters, the assumption of Rutherford trajectories for the colliding nuclei is not valid, and this is the reason that calculations have been carried out in the interval $E_{\rm c.m.}=156.7-185.8$ MeV, that is, two points less than in the experiment.

The TD-CDFT results for the PLF proton and neutron distributions are shown in Fig.~\ref{fig:distribution_nature}. Compared with the data (Fig. 1 of the supplement to Ref.~\cite{Cook2023NaC}), the calculation reproduces the peaks of the distributions, but not the widths. The theoretical distributions exhibit considerably smaller nucleon-number fluctuations, and this is because TD-DFT includes only one-body dynamics and, therefore, describes only the most probable reactions by propagating individual nucleons independently in self-consistent mean-field potentials. This means that the TD-DFT results, as well as the relation $L=(8\ln 2)~\Delta N^2$ \cite{Klich2006JPA} that is based on the assumption that the nuclear wave function is a simple Slater determinant at all times, provide a lower limit on nucleon-number fluctuation and entanglement. The experimental fluctuation of the number of protons and neutrons in the reflected nuclei and, therefore, the entanglement between the PLF and TLF are considerably larger than the TD-DFT estimate. Obviously, to reproduce the level of entanglement observed in the experiment, it will be necessary to extend the TD-DFT-based mean-field model to include quantum fluctuations. Such an extension has been
developed in our recent articles \cite{Li2023PRC_Gd,Li2024FOP}, and will be applied to multi-nucleon transfer reactions in a  forthcoming study.

We also note that, while in the experiment \cite{Cook2023NaC} the neutron-number fluctuations in the PLF are much more pronounced than the proton ones, they are not so different for the theoretical distributions in Fig.~\ref{fig:distribution_nature}. This result points to additional many-body correlations, not included in the present implementation of the TD-CDFT.
As shown in Fig.~\ref{fig:linear_regression2}, perfect linear relations between the particle-number fluctuations and entanglement are obtained by the simple linear regression. Eq.~(\ref{inequility}) is satisfied in these $^{40}$Ca+$^{208}$Pb reactions. Using Eqs.~(\ref{SV}) and~(\ref{particle_fluctuation}), the eigenvalues $d_i$ of the overlap matrix can be evaluated. Most of them are obtained either in the interval $0.04-0.05$ or $0.94-0.96$. This means that most single-nucleon wave functions are localized in the projectile-like subspace or target-like subspace, and the entanglement between the two emerging fragments is determined by the non-locality of only a few single-nucleon wave functions.

\begin{figure}[!htbp]
\centering
\includegraphics[width=0.75\textwidth]{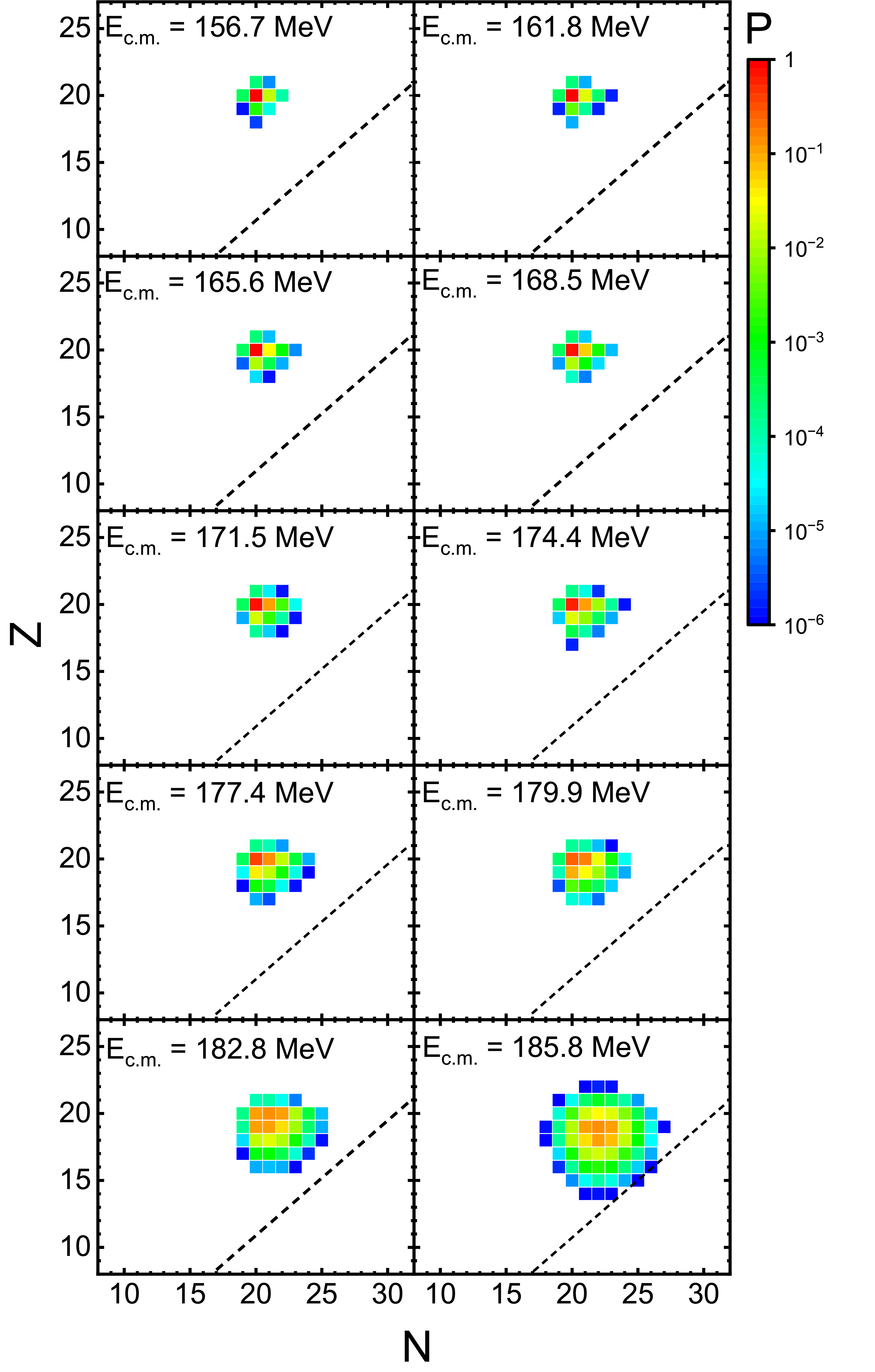}
\caption{(Color online) Proton $Z$ and neutron $N$ number distributions of the PLF for the $^{40}$Ca+$^{208}$Pb reaction, at the laboratory angle $\theta_{\rm lab}= 115^{\rm o}$ and center-of-mass energies $E_{\rm c.m.}$ from $156.7$ to $185.8$ MeV. The color scale indicates the probability of forming each nucleus. The diagonal line corresponds the isospin asymmetry equal to that of the compound nucleus $^{248}$No ($N/Z = 1.43$).  }
 \label{fig:distribution_nature}
\end{figure}

\begin{figure}[!htbp]
\centering
\includegraphics[width=0.5\textwidth]{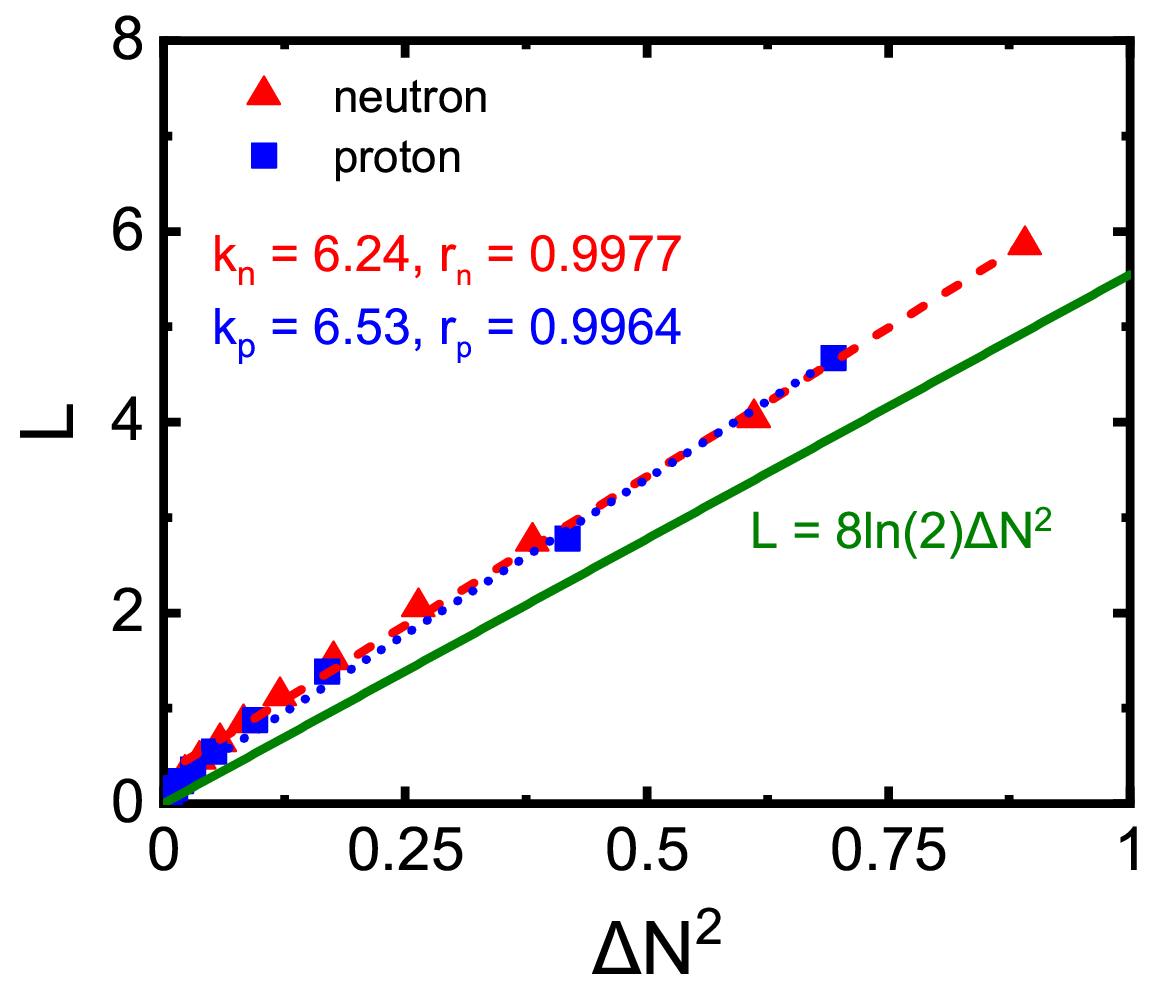}
\caption{(Color online) Same as in the caption to Fig.~\ref{fig:linear_regression}, but for the reaction $^{40}$Ca+$^{208}$Pb at laboratory angle $\theta_{\rm lab}= 115^{\rm o}$
and center-of-mass energies $E_{\rm c.m.}=156.7-185.8$ MeV. }
 \label{fig:linear_regression2}
\end{figure}

\begin{figure}[!htbp]
\centering
\includegraphics[width=0.8\textwidth]{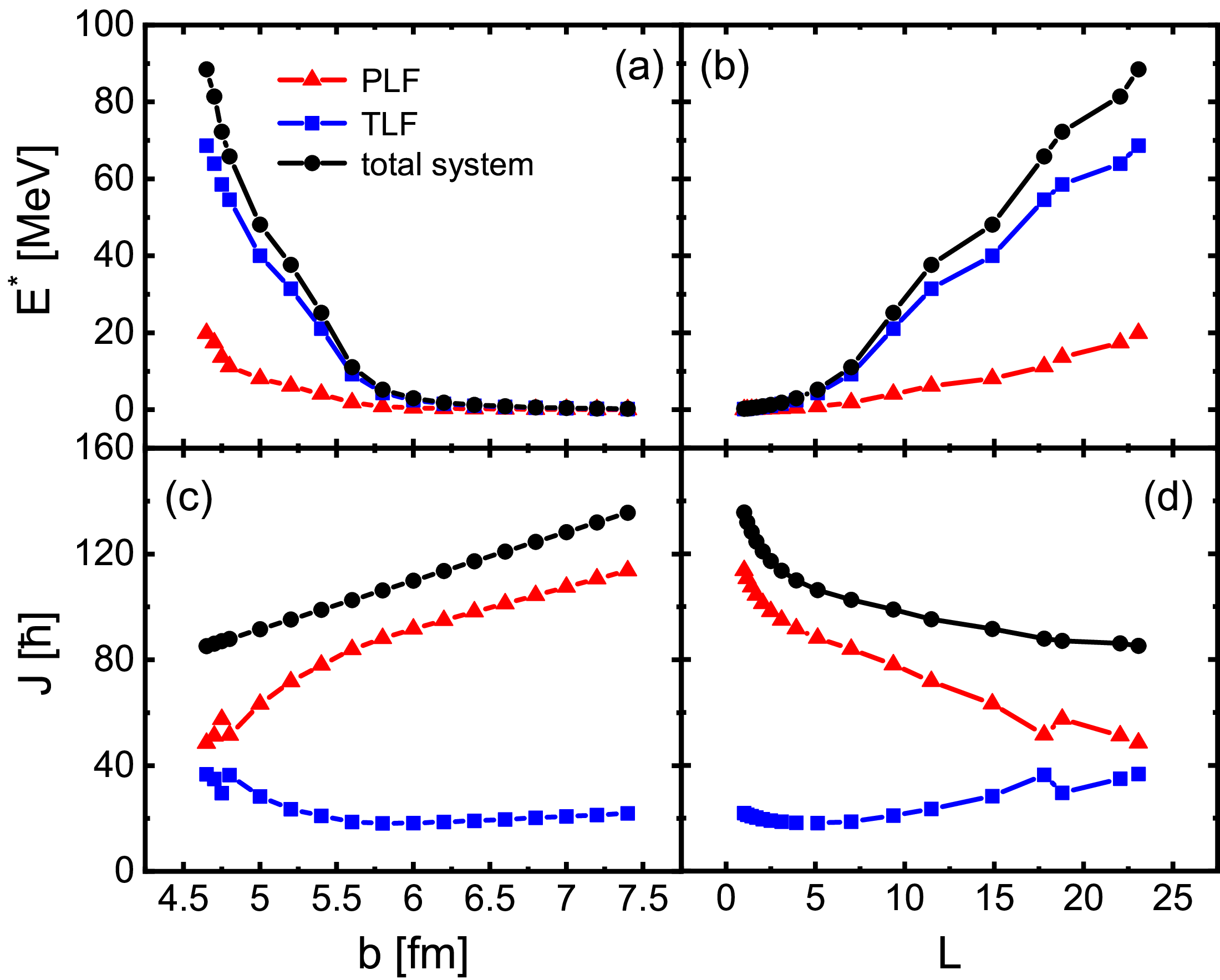}
\caption{(Color online) $^{40}$Ca+$^{208}$Pb at $E_{\rm lab}=249$ MeV. Left panel: Excitation energy (a), and angular momentum (c) of the PLF (red), TLF (blue), and the total system (black), as functions of impact parameter. The same quantities are shown in the right panels (b) and (d), as functions of the entanglement between the PLF and TLF.}
 \label{fig:Excitation_energy}
\end{figure}

The excitation energies of the PLF, TLF, and total system after the collision, are displayed in panel (a) of Fig.~\ref{fig:Excitation_energy}, as functions of the impact parameter, and in panel (b) as functions of the entanglement between the PLF and TLF. The total excitation energy, as well as the excitation energy of the TLF, exhibit a steep increase for $b \leq 5.5$ fm. At $b=4.65~{\rm fm}$, the average total excitation energy is $\approx 90$ MeV, and most of this energy $\approx 70$ MeV, resides in the TLF. The excitation energy of the PLF which is, of course, much lighter, increases more gradually, and the largest value is $\approx 20$ MeV at $b=4.65~{\rm fm}$. The panel on the right illustrates the increase of the excitation energies as functions of the entanglement. In panel (c) of Fig.~\ref{fig:Excitation_energy} we plot the expectation values of the total angular momentum, the angular momenta of the PLF and TLF, as functions of the impact parameter. The same quantities are plotted in panel (d), as functions of the entanglement between the PLF and TLF.
The total angular momentum is linearly proportional to the impact parameter, and varies between $\approx 80~\hbar$ at $b=4.65~{\rm fm}$, and $\approx 140~\hbar$ at $b=7.40~{\rm fm}$. Of course, in contrast to the excitation energy in panel (a), most of the angular momentum is carried by the PLF, except for small values of the impact parameter ($b \leq 5$ fm) at which quasifission occurs. We note that, for this typical example of a MNT reaction, in panels (b) and (d), the entanglement between the PLF and TLF is directly related to potential observables, that is, the excitation energies and angular momenta of the emerging fragments.

\begin{figure}[!htbp]
\centering
\includegraphics[width=1\textwidth]{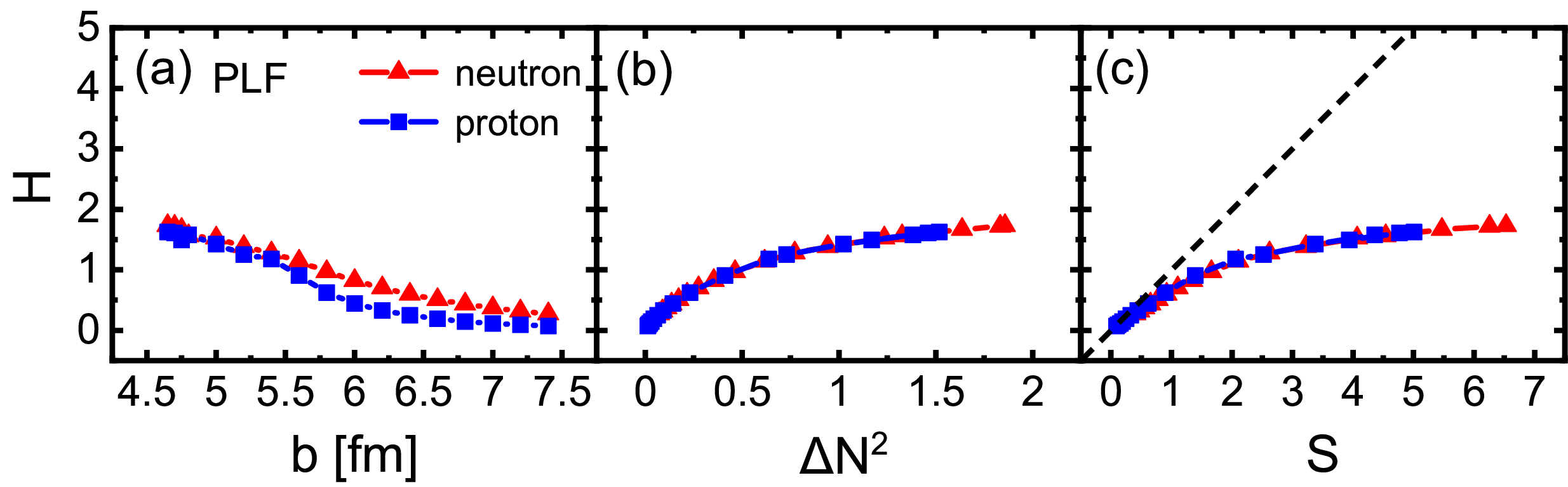}
\caption{(Color online) $^{40}$Ca+$^{208}$Pb reaction at $E_{\rm lab}=249$ MeV. Shannon entropy of the PLF for neutrons (red) and protons (blue), as a function of: the impact parameter (a), particle number fluctuations of the PLF (b), and the von Neumann entropy of the PLF (c).}
 \label{fig:Shannon_entropy}
\end{figure}
The Shannon entropy $H$ of the PLF or TLF, as defined by Eq.~(\ref{Shannon_entropy}), is computed directly from the nucleon-number probability distribution of the fragment emerging from collision. The latter is calculated by performing particle-number projection (PNP) on the fragment wave function. $H$ does not provide information on the actual number of nucleons in a fragment, rather it is a measure of the variability of the distribution of nucleons. For the PLF, in Fig.~\ref{fig:Shannon_entropy} we display the Shannon entropy for neutrons and protons as a function of the impact parameter (a), particle-number fluctuations (b), and the von Neumann entropy of the PLF. It appears that, compared to the von Neumann entropy, the Shannon entropy increases more gradually for smaller values of the impact parameter, and the dependence on the particle-number fluctuation is parabolic rather than linear. This is also clearly shown in panel (c), in which $H$, determined by the nucleon-number distribution, is always smaller than $S$, which is calculated from the reduced density matrix of the fragment. This is because the von Neumann entropy $S$ provides a measure of the total uncertainty, or lack of information about all possible observables that can be determined from the one-body reduced density matrix of the subsystem, while the Shannon entropy $H$ is a measure of the variability of a specific observable, here the nucleon number. In fact, the difference between $S$ and $H$ could be an indicator of the lack of information about the other observables that are not specifically determined (angular momentum, excitation energy).

\begin{figure}[!htbp]

\includegraphics[width=1\textwidth]{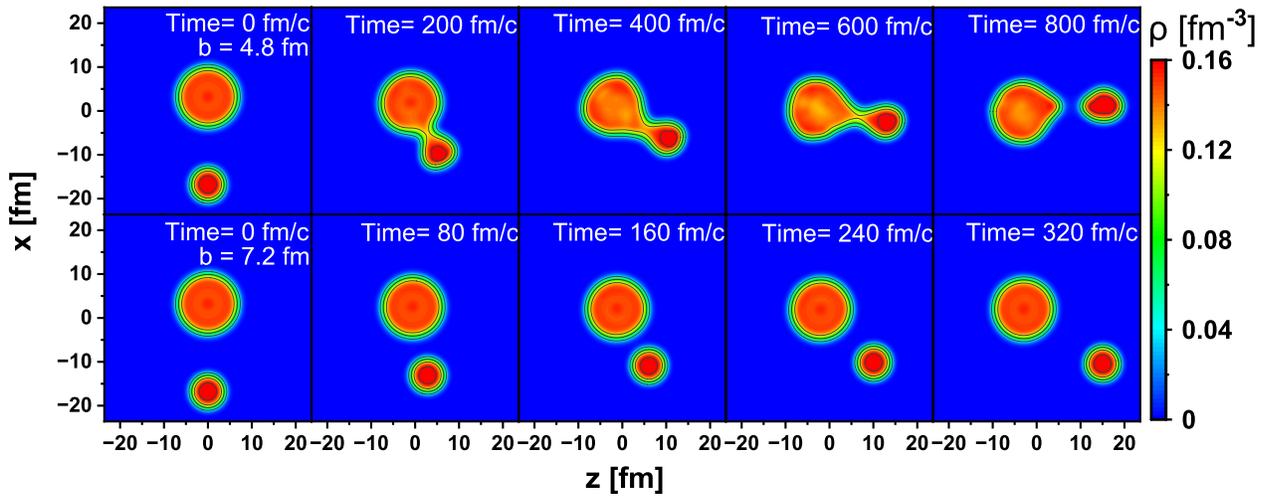}
\caption{(Color online) Snapshots of the density distribution for the $^{40}$Ca+$^{208}$Pb reaction at $E_{\rm lab}=249$ MeV. In the top panel the time-evolution is illustrated for the impact parameter $b= 4.8$ fm, while the bottom panel displays the evolution of densities for the impact parameter $b= 7.2$ fm.}
 \label{fig:density_evolution}
\end{figure}

\begin{figure}[!htbp]

\includegraphics[width=0.8\textwidth]{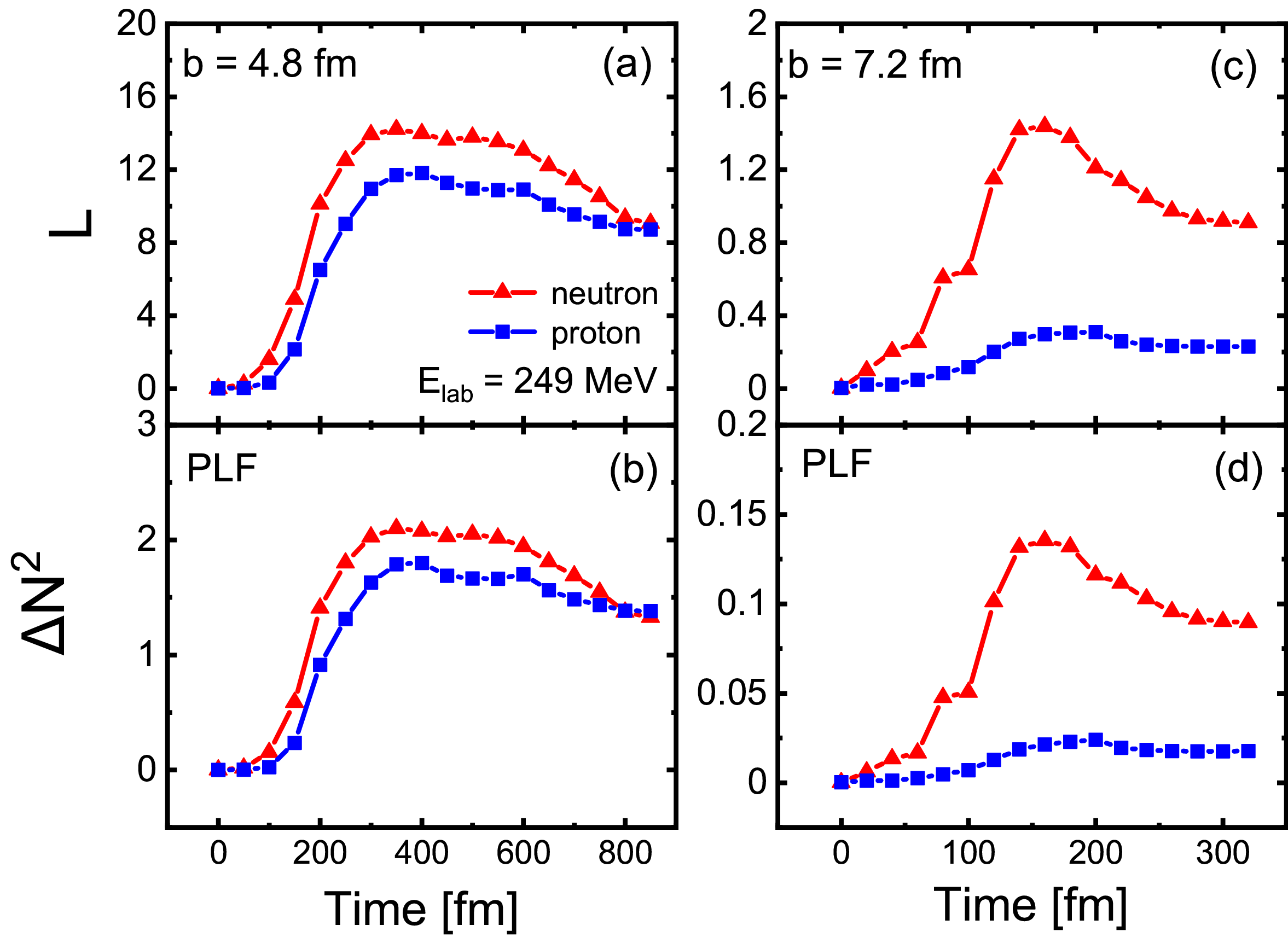}
\caption{(Color online) $^{40}$Ca+$^{208}$Pb reaction at $E_{\rm lab}=249$ MeV. Left panel: time-evolution of the entanglement between the PLF and TLF (a), particle number
 fluctuations of the PLF (b), at impact parameter $b=4.8$ fm. Right panel: same as in the panel on the left, but for the impact parameter $b= 7.2$ fm.
 Note that the right panels may not use the same vertical scale as the left ones.
 }
 \label{fig:time_evolution}
\end{figure}

It is also interesting to follow the time-evolution of the entropies, entanglement, and particle-number fluctuations. To illustrate how these quantities evolve in time, we consider two characteristic impact parameters for the reaction $^{40}$Ca+$^{208}$Pb at $E_{\rm lab}=249$ MeV. In Fig.~\ref{fig:density_evolution} several snapshots of the density distribution are shown for two values of the impact parameter: $b=4.8$ fm and $b=7.2$ fm. For the former, a di-nuclear systems is formed after $\approx 200$ fm/c, and it takes another $600$ fm/c before the PLF and TLF separate. However, during the entire interval in which the two nuclei exchange particles, it is possible to divide the space into a PLF subspace and a TLF subspace, as described in the previous section. At the impact parameter $b=7.2$ fm, virtually no nucleons are exchanged (cf. Fig.~\ref{fig:entanglement}), and the two colliding nuclei do not overlap at any time, even though they come very close to each other at about $160$ fm/c.

The corresponding time-evolution of the entanglement and nucleon-number fluctuation, is illustrated in Fig.~\ref{fig:time_evolution}.
The panels on the left and right display the results obtained for $b=4.8$ fm and $b=7.2$ fm, respectively. Since the total system is in a pure state, which is also an eigenstate of nucleon number operator, the total von Neumann entropy and nucleon-number fluctuations are zero at all times. The entanglement between the PLF and TLF, as well as the nucleon-number fluctuations of the fragments, exhibits a very interesting time evolution. For $b=4.8$ fm, both quantities steeply increase as the di-nuclear system is formed. Both for neutrons and protons the maximum occurs at $\approx 300$ fm/c, and then $L$ and $\Delta {\rm N}^2$ gradually decrease as the two fragments begin to separate in quasifission. For the larger impact parameter $b=7.2$ fm, the neutron and proton entanglements increase towards the point of closest approach, after which they gradually decrease. The nucleon-number fluctuations display the same behavior. However, when compared to the values calculated at $b=4.8$ fm, both the entanglement and the nucleon-number fluctuations are much smaller. In particular, a qualitative difference is the behavior of the proton entanglement and number fluctuation, which are, in this case, much smaller than those of the neutron distribution. This is because at $b=7.2$ fm the Coulomb repulsion dominates the reaction, and prevents the entanglement between protons in the projectile and target.

\begin{figure}[!htbp]
\centering
\includegraphics[width=0.5\textwidth]{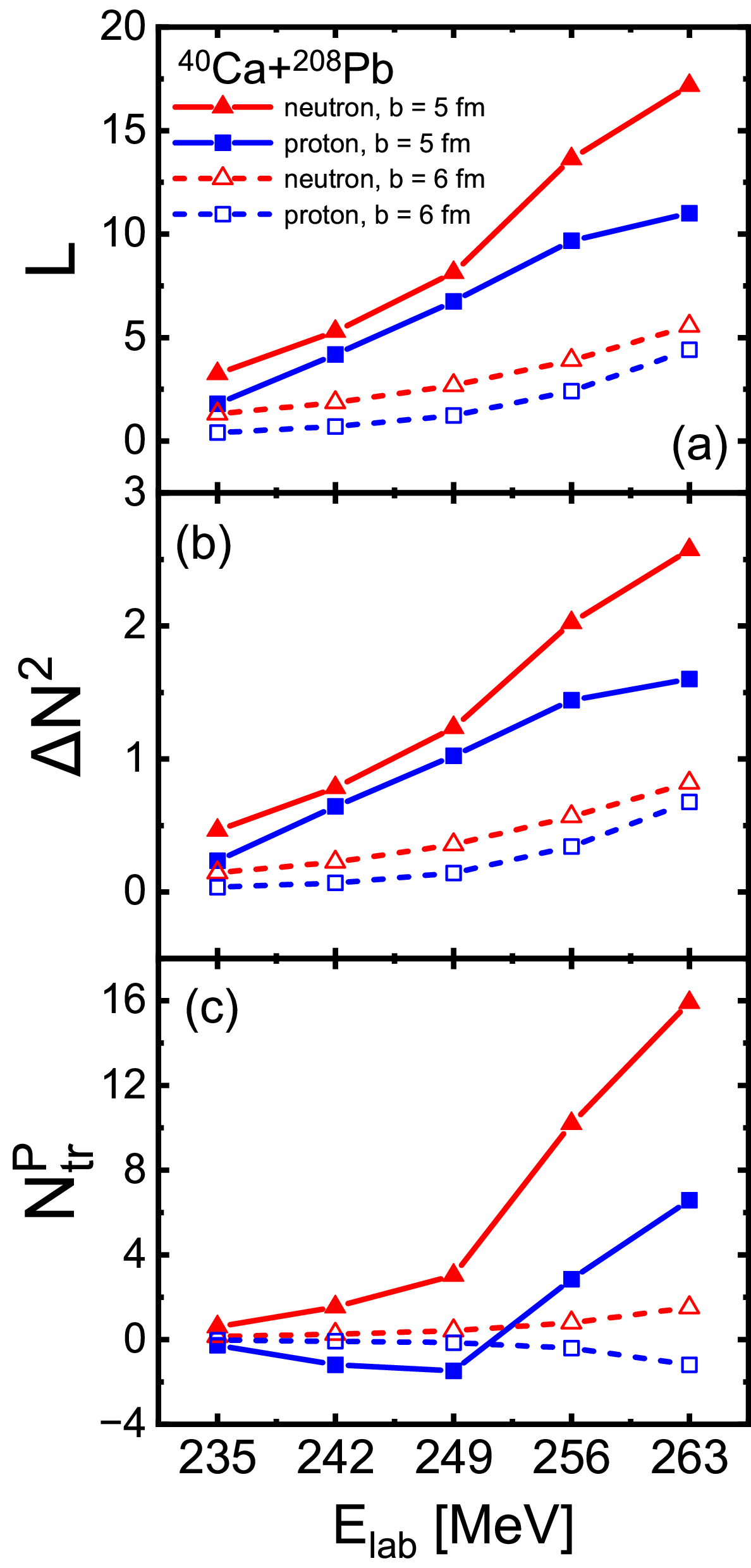}
\caption{(Color online) Entanglement between the PLF and TLF (a), particle number fluctuations of the PLF (b), and average number of transferred nucleons from the target to the projectile (c), as functions of $E_{\rm lab}$ at the impact parameter $b=5$ fm (solid), and $b=6$ fm (dashed).}
 \label{fig:incident_energy}
\end{figure}

Finally, we have also compared the entanglement and the nucleon-number fluctuations as functions of the collision energy, for two values of the impact parameter: $b=5$ fm and $b=6$ fm. Figure~\ref{fig:incident_energy} displays the entanglement after the PLF and TLF have separated (panel (a)), the number fluctuation (panel (b)), and average number of transferred nucleons from the target to the projectile (panel (c)) for a range of incident energies $E_{\rm lab}=235-263$ MeV. We find a positive correlation of the entanglement with incident energy, which means that a stronger interaction between the projectile and target leads to a more pronounced entanglement. As in the previous case, at smaller impact parameter at which quasifission dominates, the entanglement is larger and increases more steeply with incident energy. When $E_{\rm lab}$ is changing from 249 MeV to 256 MeV and $b = 5$ fm, the average number of transferred protons first declines and then increases, i.e., from $-1$ to 0, then to 2. However, the corresponding fluctuation of the proton number keeps increasing from 1 to 1.5. This indicates that the fluctuation of the proton number is not always correlated with the average number of transfer protons, in particular for the quasifission process. A much more gradual increase of the entanglement between the PLF and TLF, the nucleon-number fluctuation, and transfer number of transferred nucleons, is found at $b = 6$ fm, at which quasifission does not contribute.
\section{Summary and outlook}\label{sec_sum}

We have employed the time-dependent covariant nuclear density functional theory (TD-CDFT) to explore entanglement in multinucleon transfer reactions (MNTs). As shown in our recent analysis \cite{Zhang2024PRC_MNT}, models based on this microscopic framework can successfully be applied to the description of reaction dynamics, and reproduce the experimental cross sections for various transfer reaction channels. Here we have considered a specific example from that study, namely the reaction $^{40}$Ca $+$ $^{208}$Pb at $E_{\rm lab} = 249$ MeV, and employed the relativistic density functional PC-PK1 to compute the von Neumann entropies, entanglement between fragments, nucleon-number fluctuations, and Shannon entropy for the nucleon-number observable, as functions of the impact parameter.

From these calculations, a perfect linear correlation between the entanglement of the particle-like and target-like fragments (PLF and TLF), and the corresponding nucleon-number fluctuations has been established. The relation between the entanglement and nucleon-number fluctuation satisfies the inequality of Eq.~(\ref{inequility}), which thus provides a lower bound for the entanglement entropy, expressed in terms of the observable nucleon-number fluctuation. We have also analyzed the dependence of the entanglement between the PLF and TLF, on the corresponding excitation energies and angular momenta, as functions of the impact parameter of the reaction.

The von Neumann entropies provide information about the entanglement of states of the subsystems, but do not involve a computation of any specific observable. The measurement entropy of an observable can be expressed in terms of the Shannon entropy. For the observable number of nucleons in a fragment, we have determined the corresponding probability distributions by particle number projection, and computed the Shannon entropy as function of the impact parameter. This quantity does not depend on the actual number of nucleons in a fragment, it rather quantifies the variability of the distributions of nucleons. It has been shown that the Shannon entropy displays a parabolic dependence on particle-number fluctuation, and is always smaller than the corresponding von Neumann entropy.

Nuclear reactions provide a unique opportunity to study the time-evolution of entanglement between interacting quantum systems. Therefore, for two characteristic values of the initial impact parameter, we have analyzed the time-dependence of the entanglement and the nucleon-number fluctuations. Since the system $^{40}$Ca $+$ $^{208}$Pb is initially in a pure state that is also an eigenstate of the particle-number operator, the total entropy and number fluctuation are zero at all times. The time-evolution of the reduced density matrices of the subsystems is, however, not unitary and the entanglement (nucleon-number fluctuation) exhibits an interesting time-dependence. As the projectile approaches the target, the entanglement sharply increases and reaches a maximum at a distance of closest approach, or maximal overlap when a di-nuclear system is formed. As the PLF and TLF begin to separate, the entanglement gradually decreases, eventually reaching the asymptotic value for the final subsystems. Finally, an interesting quantitative result has also been obtained for the positive correlation between the entanglement (nucleon-number fluctuation) and collision energy, in an interval of incident energies $E_{\rm lab}=235-263$ MeV.

This study opens many attractive avenues for the exploration of quantum entanglement in time-dependent processes. A straightforward extension of the present work is a study of entanglement in fission. In general, the initial state will not be a pure state, and the initial entropy will depend on the excitation energy of the fissioning nucleus. The entanglement of the fission fragments can be analyzed as a function of the charge (mass) yields distribution, kinetic energy distribution, and angular momenta. In addition to the entanglement of states (von Neumann entropies), which is an observable-independent measure of the correlations between fission fragments, the entanglement of specific observables can be quantified by computing the corresponding Shannon entropies. For MNT reactions, one can trivially extend the present study to cases in which more than two fragments emerge from the collision such as, for instance, the case of ternary quasifission.

\begin{acknowledgments}
This work has been supported in part by the High-End Foreign Experts Plan of China,
National Key Research and Development Program of China (Contract No.2018YFA0404400),
the National Natural Science Foundation of China (Grants No.12070131001, No.11935003, No.11975031,and No.12141501),
the High-Performance Computing Platform of Peking University,
by the project ``Implementation of cutting-edge research and its application as part of the Scientific Center of Excellence for Quantum and Complex Systems, and Representations of Lie Algebras'', PK.1.1.02, European Union, European Regional Development Fund, and by the Croatian Science Foundation under the project Relativistic Nuclear Many-Body Theory in the Multimessenger Observation Era (IP-2022-10-7773).
\end{acknowledgments}

\bigskip


%

\end{CJK*}
\end{document}